\documentclass{evn2002}
\begin{document}
   \title{The Diverse Properties of GPS Sources}
          
   \author{ M. L. Lister\inst{1}, K. I. Kellermann\inst{1},
          \and
         I. I. K. Pauliny-Toth\inst{2}
          }

   \institute{National Radio Astronomy Observatory, 520 Edgemont Road,
Charlottesville VA 22901, USA
         \and
             Max-Planck Institute fur Radioastronomie, Auf dem Huegel
69, D-53121 Bonn 1, Germany}

\abstract{ We discuss the morphology and kinematics of five
gigahertz-peaked spectrum (GPS) sources that have been observed with
the VLBA.  We find a wide range of observed properties including
core-jet structure, superluminal motion, variability, extended
structure, and polarization, all of which appear to deviate from
commonly-accepted GPS paradigms (e.g., O'Dea \cite{OD98}). We suggest
that the observed low frequency cutoff in GPS sources may be primarily
due to free-free absorption rather than synchrotron self-absorption.

}


   \maketitle
%

\section{Introduction}
GPS sources are characterized by their sharp low-frequency spectral
cutoff and the absence of large scale structure.  Previous
observations have suggested a simple double morphology with no
evidence for significant relative component motions.  In this paper,
we present new VLBA observations of five well-known GPS sources whose
diverse properties are difficult to understand in terms of simple
conventional models where the GPS sources are classical self-absorbed
synchrotron sources that are the precursors of double-lobed
radio galaxies.

\section{Individual GPS sources}
\subsection{CTA 102 (2230+114)}
The quasar CTA~102 (z = 1.04) was one of the first radio sources found
to have a pronounced cutoff at low frequencies (Kellermann et
al. \cite{KLA62}). This was later interpreted by Slysh (\cite{S63})
and Williams (\cite{W66}) as the result of synchrotron self-absorption
(SSA),
and provided the first evidence for highly compact structure in
extragalactic radio sources. Observations by Sholomitsky (\cite{S65})
showed remarkable variations in the decimeter wavelength flux density
of CTA 102 which were difficult to understand in the framework of conventional
synchrotron theory.  Subsequently, many other quasars and a few AGN
with flat radio spectra were found to show similar variability, but
the discovery of superluminal motion in many quasars appeared to
provide a simple interpretation in terms of relativistic beaming

Recent multi-wavelength observations of CTA 102 have continued to
show large variations in flux density (Aller and Aller, private
communication).  Indeed, in recent years the spectrum is no longer
recognizable as a GPS source, and VLBA observations made at 2 cm since
1995 show a characteristic core-jet structure which we typically associate
with flat spectrum radio sources (Fig. 1).  

In spite of CTA~102's rapid flux density variability, our multi-epoch
VLBA observations show no evidence of any component motions.  It is
possible that the relativistic fluid is moving smoothly along the jet,
and is not reflected by observable pattern motions described by the
bright features seen in Fig. 1.  Rather, these features may merely
mark apparent brightness enhancements at the location of each bend, as
would be expected from a relativistic helical jet at locations where
the fluid velocity vector is directed at the observer.

\begin{figure}
\centering
\vspace{340pt}
\includegraphics{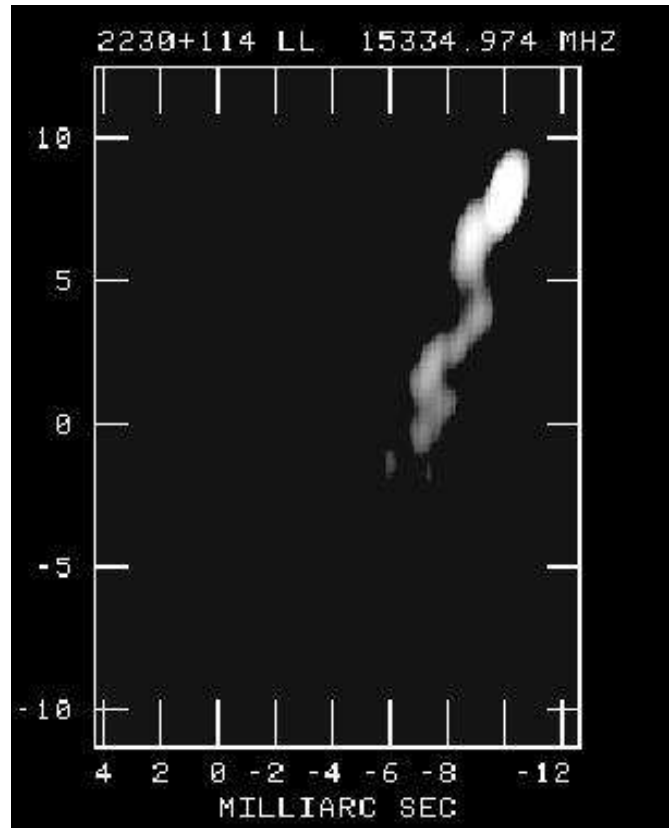}

\caption{VLBA 2cm image of CTA 102 at epoch 1996 October. The
greyscale indicates total intensity.} 
\end{figure}

\begin{figure*}
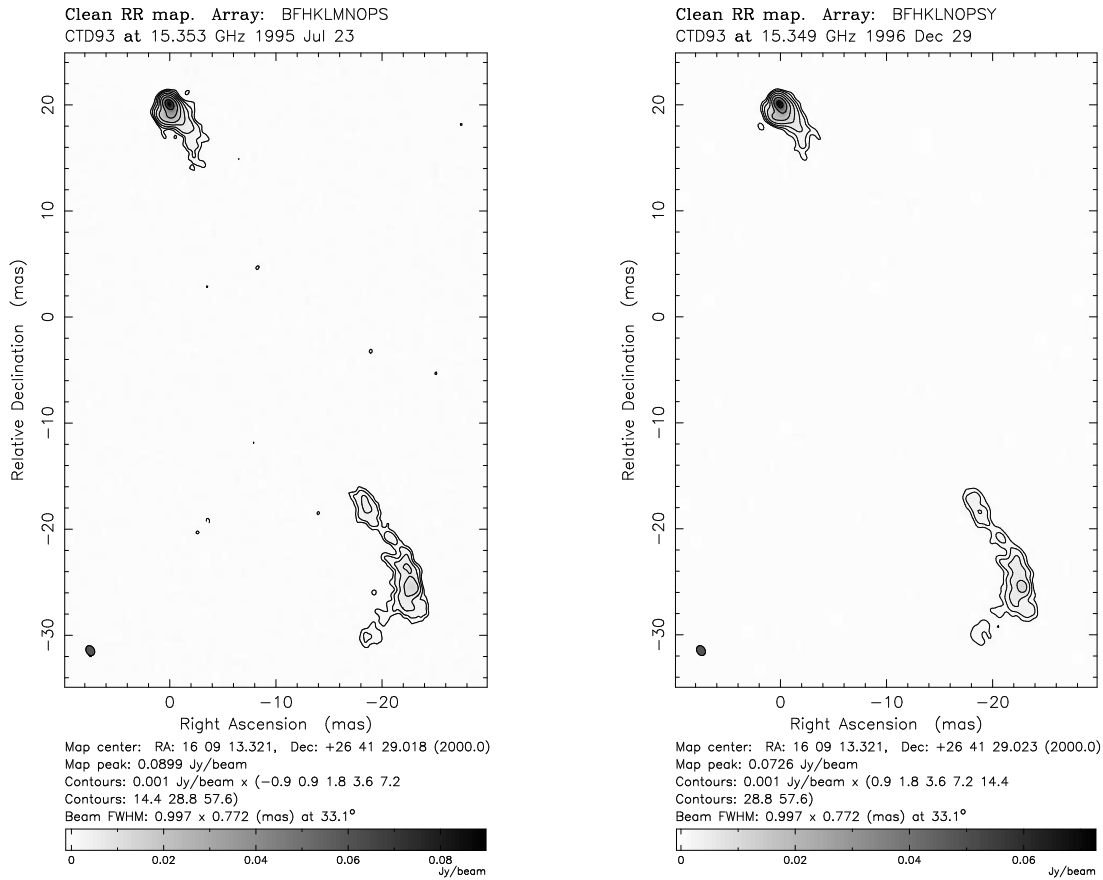

\centering
\vspace{350pt}
\includegraphics{1995.ps}
\includegraphics{1996.ps}

\caption{VLBA images at 2 cm of CTD~93 at epochs 1995 July (left) and
1996 Dec. (right) drawn with the same contour levels.
         \label{ctd93} } 
	
\end{figure*}

\subsection{CTD~93 (1607+268)}

CTD~93 is an AGN identified with a $m_V = 20.3$ galaxy at $z=
0.47$. With a radio luminosity of $3 \times 10^{27}\; \rm W \;
Hz^{-1}$, it is one of the brightest radio galaxies known.  It was
first discovered in a 20 cm survey (Kellermann \& Read \cite{KR63})
and was later found to be a GPS source by Kellermann
(\cite{K66}). Early VLBI observations (Phillips \& Mutel \cite{PM80},
Phillips \& Shaffer \cite{PS83}) indicated an apparently simple
symmetric double structure with an unusually large component
separation.  Subsequent VLBI observations of other GPS sources led to
the paradigm which associates GPS sources with symmetric double
structure rather than classical core-jet structure (Phillips \& Mutel
\cite{PM82}).  However, higher-resolution VLBA observations of CTD~93
made in July 1995 (Fig.~\ref{ctd93}) revealed the southern component
was in fact a narrow, sharply bent jet, while the northern component
was a compact feature likely to be the core (Shaffer, Kellermann, \&
Cornwell \cite{SKC99}).  Multi-frequency observations showed,
surprisingly, that the southern low-surface-brightness jet has a
similar spectrum as the presumed core component.  This is contrary to
what would be expected if the low frequency spectral cutoff is a
result of SSA, and suggests that the low frequency cutoff may be the
result of free-free absorption (FFA) from a surrounding ionized
medium.  We have re-observed CTD~93 with the VLBA on 29 Dec 1996, and
find identical structure to the earlier observations, with no evidence
for any change in component separation during the 18 month period
between the two epochs.  Our observations place an upper limit to the
relative component motion of 0.5c.

\subsection{2134+004}

This quasar is similar to CTD~93 in that it was one of the
earliest-recognized GPS sources (Shimmins et al.  \cite {SS67}), and is
exceedingly luminous at centimeter wavelengths. It is identified with
a $m_V = 18$ quasar at z=1.93 and  is highly variable in the optical
(Gottlieb \& Liller \cite{GL78}), with occasional flarings
of more than three magnitudes. Its flux density is relatively
constant, however, at radio
wavelengths (Aller et al.
\cite{AAL85})\footnote{http://www.astro.lsa.umich.edu/obs/radiotel/umrao.html}.
Like CTD~93, it was also thought to be a symmetric double, but our
multi-wavelength VLBA observations indicate that the more compact
easternmost component is associated with the base of a one-sided
jet. There is a faint bridge that reaches toward the western component
which we associate with highly beamed radiation from a portion of
a highly curved jet that is directed toward the observer.
Multi-epoch VLBA observations at 2 cm provide marginal evidence that
the two components are drawing closer in time. However, our fitted
core positions may be affected by blending of components below the
resolution level of our images (see, e.g., Lister and Smith
\cite{LS00}).

   \begin{figure}
   \centering
   \vspace{230pt}
   \includegraphics{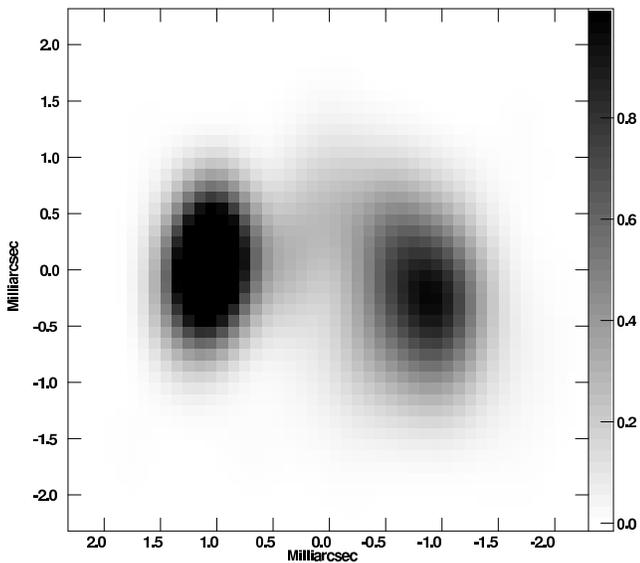}
\caption{VLBA 2cm image of 2134+004 at epoch 1995 May.   \label{2134+004} }  

   \end{figure}

\subsection{OQ~208 (1404+286)}

\begin{figure}
   \centering
   \vspace{230pt}
   \includegraphics{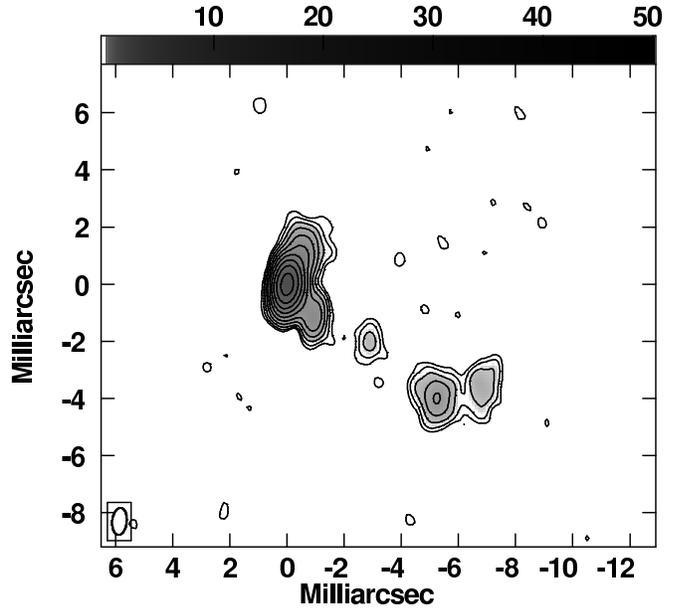}
\caption{VLBA 2cm image of OQ~208 at epoch 1997 Aug. The total
contours levels are 1, 2, 4, 8, 16, 32, 64, 128, 256, and 512 $\rm
mJy\; beam^{-1}$.   \label{OQ208} }  

   \end{figure}

OQ~208 is one of the closest known GPS sources (z = 0.08), and is
associated with the broad line radio galaxy Mkn 668.  Its radio
morphology is characterized by a very weak central core and two
sharply bent jets (Fig. \ref{OQ208}). The core appears to have a
variable flux density, and is evident at only some of our VLBA epochs.
The eastern jet undergoes a ninety degree bend similar to the southern
component of CTD~93. There is no apparent motion of the jet components
with respect to the presumed core with an upper limit of about 0.5c.

\begin{figure}
\centering
\vspace{468pt}
\includegraphics{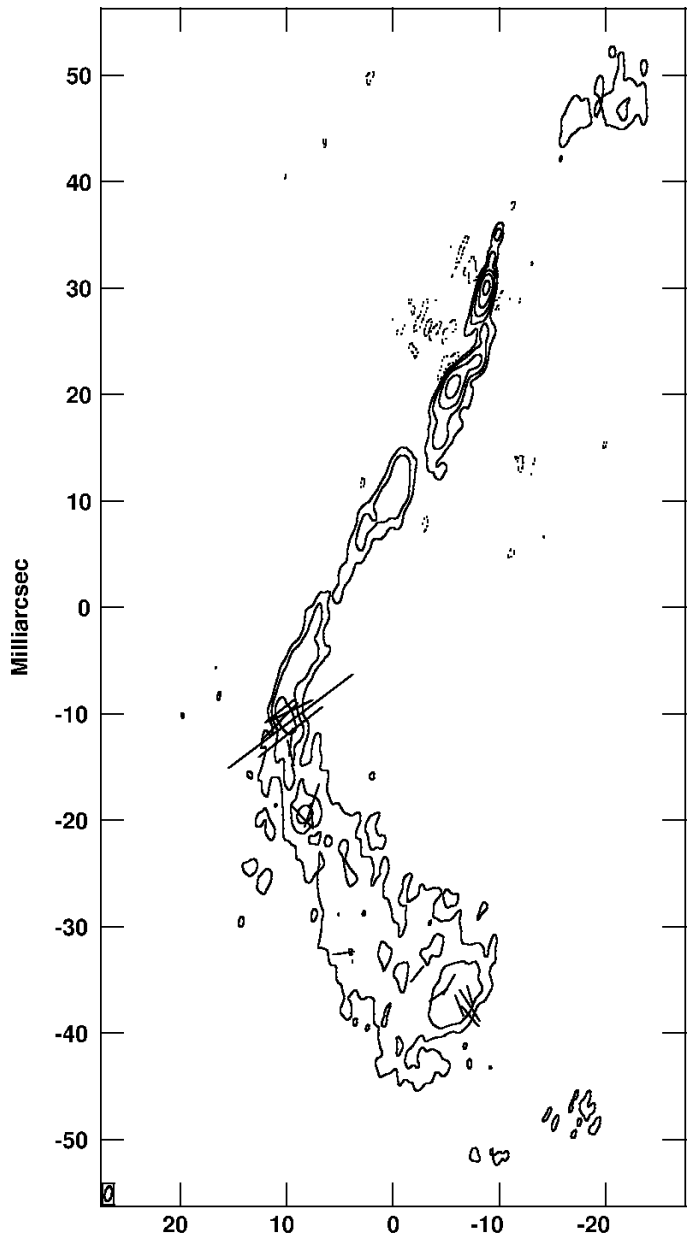}
\caption{2 cm VLBA image of PKS 1345+125 at epoch  4  Jan 2001,
with electric vectors superimposed. The lowest I contour is $\pm 0.4\; \rm
mJy\; beam^{-1}$, with successive contours increasing by factors of 4.
\label{1345+125} }
\end{figure}

\subsection{PKS 1345+125 (4C +12.50)}

The host galaxy of PKS 1345+125 is an ULIRG at z=0.12 that is rich in
both molecular and atomic gas (Evans et al. \cite{EKM99}, Mirabel
\cite{M89}).  The host contains two optical nuclei separated by $\sim 3.5$
kpc that appear to have undergone a recent merger, as evidenced by its
tidal tail structure and distorted isophotes (Gilmore \& Shaw
\cite{GS86}). The western nucleus harbors an obscured quasar and a
powerful compact radio source with a GPS spectrum and a sharp spectral
cutoff near 400 MHz. VLBA observations reveal a well-defined jet
structure with a total extent of $\sim 100$ mas = 220 pc
(Fig.~\ref{1345+125}). The brightest feature in our full-track VLBA
image has an inverted spectrum and was identified as the core by
Stanghellini et al. (\cite{SDO01}). Although the core is weakly
polarized ($m = 0.3\%$), the fractional polarization reaches very high
levels in the southern jet, with $m = 30\%$ at the major bend and $m =
60\%$ at the very tip of the jet. The electric vectors in both regions
are perpendicular to the total intensity contours, which suggests that
shocks in the flow have ordered the magnetic field of the jet. 

The presence of a relatively bright northern counter-jet and mild superluminal
motion (1.2c) in the southern jet (Lister et al. \cite{LKV02}) imply
that the jet axis of this source lies fairly close to the plane of the
sky. The overall bent morphology suggests that the jet nozzle
direction may be slowly changing with time, perhaps as a result of the
merger event. Lister et al. (\cite{LKV02}) were able to obtain a good
fit to both the ridge lines and brightness distributions of the jet
and counter-jet using a simple precessing jet model. The viewing angle
to the inner jet in their best-fit model (62 deg.) is consistent with
the angle inferred from the observed superluminal speed and
jet/counter-jet flux density ratio.

\section{Summary}

All five of the sources we have studied have sharply bent jet
structure. We find no other systematic properties which distinguish
these GPS sources other than their peaked radio spectrum, which may be
due to a combination of SSA and FFA from a surrounding ionized medium.
CTA~102, 2134+004, and CTD~93 have bright cores plus
fainter bent jets typical of what is observed in other quasars and
AGN, whereas PKS 1345+125 and OQ~208 have double-sided jet structures that
end in sharp bends.   The identically-peaked spectra of the
separate components of individual GPS sources implies little
differential Doppler shift of the spectral components if the cutoff is
from SSA. This interpretation is supported by the absence of any
differential component motions, possibly because all the components we
are observing are moving at the same velocity with respect to an
unidentified core below our level of detection.  But this
interpretation is not consistent with observed the highly bent structure
which would appear to require changes in the Lorenz factor
corresponding to changes in the direction of the flow.

The common cutoff frequency across components of widely
different surface brightness therefore suggests that some or all of the low
frequency cutoff may be due to  FFA rather than SSA.  In this case the
small angular size predicted from the early spectral observations and
the assumption of SSA would be coincidental.

\begin{acknowledgements}

This paper is based in part on work done in
collaboration with H. Aller, M.  Aller, M. Cohen, D. Homan, M. Kalder,
Y. Kovalev, D. Shaffer, R. Vermeulen, and A. Zensus.

\end{acknowledgements}

\end{document}